\documentclass[12pt]{article}

\usepackage{epsfig,amsmath,amssymb,graphics,graphicx} 

\newcommand{\be}{\begin{equation}}
\newcommand{\ee}{\end{equation}}
\newcommand{\bi}[1]{\vspace{-3mm} \bibitem{#1}}

\begin{document}

\begin{center}
{\it Journal of Physics A 39 (2006) 14895-14910}
\end{center}

\vskip 3mm

\begin{center}
{\Large \bf Continuous Limit of Discrete Systems with \\ 
Long-Range Interaction} 
\vskip 5 mm

{\large \bf Vasily E. Tarasov} \\

\vskip 3mm

{\it Skobeltsyn Institute of Nuclear Physics, \\
Moscow State University, Moscow 119991, Russia } \\
{E-mail: tarasov@theory.sinp.msu.ru}
\end{center}

\vskip 11 mm

\begin{abstract}
Discrete systems with long-range interactions are considered. 
Continuous medium models as continuous limit of discrete chain system 
are defined.  Long-range interactions of chain elements that
give the fractional equations for the medium model are discussed.
The chain equations of motion with long-range interaction
are mapped into the continuum equation
with the Riesz fractional derivative. 
We formulate the consistent definition of continuous limit 
for the systems with long-range interactions.
In this paper, we consider a wide class of long-range interactions 
that give fractional medium equations in the continuous limit.
The power-law interaction is a special case of this class.
\end{abstract}

\vskip 3 mm


%%%45.10.Hj Perturbation and fractional calculus methods
%%%45.05.+x General theory of classical mechanics of discrete systems
%%%45.50.-j Dynamics and kinematics of a particle and a system of particles
%%%05.45.Df Fractals

\vskip 3 mm


\vskip 11 mm

\section{Introduction}

Derivatives or integrals of noninteger order \cite{OS,SKM,MR,Podlubny,KST} 
have found many applications in recent studies in mechanics and physics 
\cite{Zaslavsky2,Zaslavsky1,Mainardi,Tar1,Tar2}. 
Equations which involve derivatives or integrals of non-integer order 
are very successful in describing anomalous kinetics and transport 
and continuous time random walks \cite{Ga1,Ga2,Ga3,Ga4,Ga5}.
Usually, the fractional equations for dynamics or kinetics 
appear as some phenomenological models. 
Recently, a method to obtain fractional analogues of equations 
of motion was considered for sets of coupled particles 
with a long-range interaction \cite{LZ,TZ3,KZT,KZ}. 
Examples of systems with interacting oscillators, spins or waves 
are used for many applications in physics, chemistry, biology
\cite{Dyson,J,CMP,NakTak,S,Kur,Ruf,BK,Ish,BKZ,Br6,PV,Br4,GF,LLI}. 
In the continuous limit, the equations of motion for discrete systems
give the continuous medium equation. 
The procedure has already been used 
to derive fractional sine-Gordon and fractional wave 
Hilbert equation \cite{LZ,KZT}, to study synchronization of 
coupled oscillators \cite{TZ3}, to derive fractional 
Ginzburg-Landau equation \cite{TZ3} and for 
chaos in discrete nonlinear Schr\"odinger equation \cite{KZ}.
In \cite{LZ,KZT,TZ3,KZ}, only the power-law long-range 
interactions are considered. 
In this paper, we consider a wide class of long-range interactions 
that give fractional medium equations in continuous limit.
The power-law interaction is a special case of this class.

Long-range interaction (LRI) has been the subject of investigations
for a long time.
An infinite one-dimensional Ising model with LRI
was considered by Dyson \cite{Dyson}.
The $d$-dimensional classical Heisenberg model with long-range 
interaction is  described in \cite{J,CMP}, and
their quantum generalization 
can be found in \cite{NakTak,S}.  
Solitons in a one-dimensional lattice with the 
long-range Lennard-Jones-type interaction were considered in \cite{Ish}. 
Kinks in the Frenkel-Kontorova model with long-range 
interparticle interactions 
were studied in \cite{BKZ}. The properties of time periodic 
spatially localized solutions (breathers) 
on discrete chains in the presence of algebraically decaying interactions 
were described in \cite{Br4,GF}. 
Energy and decay properties of discrete breathers in systems with LRI 
have also been studied in the framework of the Klein-Gordon 
\cite{BK}, and discrete nonlinear Schr\"odinger equations \cite{Br6}. 
A main property of the dynamics described by the equation with 
fractional space derivatives is that the 
solutions have power-like tails. 
The lattice models with power-like long-range 
interactions \cite{PV,Br4,GF,AEL,AK,APV,KZT} have similar properties. 
As was shown in \cite{TZ3,KZT,KZ}, 
the analysis of the equations with fractional derivatives can provide 
results for the space asymptotics of their solutions. 

The goal of this paper is to 
formulate the consistent definition of continuous limit 
(transform operation) for the systems with long-range interactions (LRI).
This aim is realized by Propositions 1, 4, and Definitions 1, 2.
The power-law LRI is considered in \cite{LZ,TZ3,KZT,KZ}. 
The exact continuous limit results for power-law LRI 
were formulated in Propositions 2, 3.
This operation is used to consider a wide class of long-range interactions 
that can be called alpha-interaction.
In continuous limit, the equations of motion gives the medium equations
with fractional derivatives.
The power-law interaction is a special case of this class 
of $\alpha$-interactions.
We show how the continuous limit for the systems 
of oscillators with long-range interaction
can be described by the corresponding 
fractional equation.

In Sec. 2, the transform operation that maps the discrete
equations into continuous medium equation is defined.
In Sec. 3, the Fourier series transform of the equations of a system 
with long-range interaction is realized.
A wide class of long-range interactions 
that can give the fractional equations in the continuous limit
is considered.
In Sec. 4, the fractional equations are obtained from
three-dimensional discrete system. 
In Sec. 5, the linear power-law long-range interactions with 
positive integer and noninteger powers are considered.
The correspondent continuous medium equations are discussed.
In Sec. 6, the nonlinear long-range interactions for
the discrete systems are used to derive the Burgers, 
Korteweg-de Vries, and Boussinesq equations 
and their fractional generalizations in the continuous limit. 
The conclusion is given in Sec. 7.

%%%%%%%%%%%%%%%%%%%%%%%%%%%%%%%%%%%%
\section{Transform operation}

Let us consider a one-dimensional system of 
interacting oscillators that are described by 
the equations of motion,
\be \label{Main_Eq}
\frac{\partial^s u_n}{\partial t^s} = g  
\hat I_{n} (u)  + F (u_n) ,
\ee
where $s=1,2$, and $u_n$ are displacements from the equilibrium. 
The terms $F(u_n)$ characterize an interaction of the oscillators   
with the external on-site force. 
The term $\hat I_{n}(u)$ is defined by
\be \label{Z3}
\hat I_{n} (u) \equiv 
\sum_{\substack{m=-\infty \\ m \ne n}}^{+\infty} \; 
J(n,m) \; W(u_n,u_m) ,
\ee
and it takes into account the interaction 
of the oscillators in the system. \\

For linear long-range interaction we have $W(u_n,u_m)=u_n-u_m$, 
and the interaction term (\ref{Z3}) is
\be \label{Z3b}
\hat I_{n} (u) \equiv 
\sum_{\substack{m=-\infty \\ m \ne n}}^{+\infty} \; 
J(n,m) \; [u_n-u_m] .
\ee
In this paper, we consider a wide class of interactions (\ref{Z3b}) 
that create a possibility of presenting the continuous medium 
equations with fractional derivatives.
We also discuss the term (\ref{Z3}) with $W(u_n,u_m)=f(u_n)-f(u_m)$ 
as nonlinear long-range interaction. 
As the examples, we consider $f(u)=u^2$ and $f(u)=u-g u^2$ 
that gives the Burgers, Korteweg-de Vries and Boussinesq equations 
and their fractional generalizations in the continuous limit. 

Let us define the operation, which transforms equations (\ref{Main_Eq}) 
for $u_n(t)$ into continuous medium equation for $u(x,t)$. 
We assume that $u_n(t)$ are Fourier coefficients
of some function $\hat{u}(k,t)$.
Then we define the field $\hat{u}(k,t)$ on $[-K/2, K/2]$ as 
\be \label{ukt}
\hat{u}(k,t) = \sum_{n=-\infty}^{+\infty} \; u_n(t) \; e^{-i k x_n} =
{\cal F}_{\Delta} \{u_n(t)\} ,
\ee
where  $x_n = n \Delta x$, $\Delta x=2\pi/K$ is 
distance between oscillators, and
\be \label{un} 
u_n(t) = \frac{1}{K} \int_{-K/2}^{+K/2} dk \ \hat{u}(k,t) \; e^{i k x_n}= 
{\cal F}^{-1}_{\Delta} \{ \hat{u}(k,t) \} . 
\ee
These equations are the basis for the Fourier transform, 
which is obtained by transforming 
from a discrete variable to a continuous one in 
the limit $\Delta x \rightarrow 0$ ($K \rightarrow \infty$). 
The Fourier transform can be derived from (\ref{ukt}) and (\ref{un}) 
in the limit as $\Delta x \rightarrow 0$.
Replace the discrete $u_n(t)=(2\pi/K) u(x_n,t)$ with continuous $u(x,t)$ 
while letting $x_n=n\Delta x= 2\pi n/K \rightarrow x$.
Then change the sum to an integral, and 
equations (\ref{ukt}), (\ref{un}) become
\be \label{ukt2} 
\tilde{u}(k,t)=\int^{+\infty}_{-\infty} dx \ e^{-ikx} u(x,t) = 
{\cal F} \{ u(x,t) \}, 
\ee
\be \label{uxt}
u(x,t)=\frac{1}{2\pi} \int^{+\infty}_{-\infty} dk \ e^{ikx} \tilde{u}(k,t) =
 {\cal F}^{-1} \{ \tilde{u}(k,t) \}. 
\ee
Here,  
\be 
\tilde{u}(k,t)= {\cal L} \hat{u}(k,t), 
\ee
and ${\cal L}$ denotes the passage 
to the limit $\Delta x \rightarrow 0$ ($K \rightarrow \infty$).
Note that $\tilde{u}(k,t)$ is a Fourier transform of the field $u(x,t)$,
and $\hat{u}(k,t)$ is a Fourier series transform of $u_n(t)$,
where we can use $u_n(t)=(2\pi/K) u(n\Delta x,t)$.
The function $\tilde{u}(k,t)$ can be derived from $\hat{u}(k,t)$
in the limit $\Delta x \rightarrow 0$.

The map of a discrete model into the continuous one 
can be defined by the transform operation. \\

{\bf Definition 1.}
{\it Transform operation $\hat T$ is a combination 
$\hat T={\cal F}^{-1} {\cal L} \ {\cal F}_{\Delta}$
of the operations: \\
1) The Fourier series transform:
\be \label{O1}
{\cal F}_{\Delta}: \quad u_n(t) \rightarrow {\cal F}_{\Delta}\{ u_n(t)\}=
\hat{u}(k,t) ;
\ee
2) The passage to the limit $\Delta x \rightarrow 0$:
\be
{\cal L}: \quad \hat{u}(k,t) \rightarrow {\cal L} \{\hat{u}(k,t)\}=
\tilde{u}(k,t) ;
\ee
3) The inverse Fourier transform: }
\be
{\cal F}^{-1}: \quad \tilde{u}(k,t) \rightarrow 
{\cal F}^{-1} \{ \tilde{u}(k,t)\}=u(x,t) .
\ee

The operation $\hat T={\cal F}^{-1} {\cal L} \ {\cal F}_{\Delta}$ 
is called a transform operation, since it
performs a transform of a discrete model of coupled oscillators
into the continuous medium model.

%%%%%%%%%%%%%%%%%%%%%%%%%%%%%%%%%%%%%%%%%%%%%%%%%%
\section{From discrete to continuous equation}

Let us consider the interparticle interaction 
that is described by (\ref{Z3b}), where $J(n,m)$ satisfies the conditions
\be \label{Jnm}
J(n,m)=J(n-m)=J(m-n) , \qquad \sum^{\infty}_{n=1} |J(n)|^2 < \infty .
\ee
Note that $J(-n)=J(n)$. \\

\noindent
{\bf Definition 2.} 
{\it The interaction term (\ref{Z3}) and (\ref{Jnm}) in the equation of 
motion  (\ref{Main_Eq}) is called $\alpha$-interaction if the function 
\be \label{Jak}
\hat{J}_{\alpha}(k)=\sum^{+\infty}_{\substack{n=-\infty \\ n\not=0}} 
e^{-ikn} J(n) = 2 \sum^{\infty}_{n=1} J(n) \cos(kn) 
\ee
satisfies the condition
\be \label{Aa}
\lim_{k \rightarrow 0} 
\frac{[\hat{J}_{\alpha}(k)- \hat{J}_{\alpha}(0)]}{|k|^{\alpha}} 
=A_{\alpha},
\ee
where $\alpha>0$ and $0<|A_{\alpha}|< \infty$.} \\

Condition (\ref{Aa}) means that 
$\hat{J}_{\alpha}(k)-\hat{J}_{\alpha}(0)=O(|k|^{\alpha})$, i.e.,  
\be \label{AR}
\hat{J}_{\alpha}(k)- \hat{J}_{\alpha}(0)=
A_{\alpha} |k|^{\alpha} +R_{\alpha}(k),
\ee
for $k\rightarrow 0$, where
\be
 \lim_{k \rightarrow 0} \ R_{\alpha}(k) / |k|^{\alpha}  =0 .
\ee

Examples of functions $J(n)$ for $\alpha$-interactions 
can be summarized in the table of the Appendix.\\

{\bf Proposition 1.}
{\it The transform operation $\hat T$ maps the discrete equations of motion
\be \label{C1}
\frac{\partial^s u_n(t)}{\partial t^s} = 
g  \sum_{\substack{m=-\infty \\ m \ne n}}^{+\infty} \; 
J(n,m) \; [u_n(t) -u_m(t)] + F (u_n(t)) 
\ee
with noninteger $\alpha$-interaction 
into the fractional continuous medium equations: 
\be \label{CME}
\frac{\partial^s}{\partial t^s} u(x,t) -
G_{\alpha} A_{\alpha} \frac{\partial^{\alpha}}{\partial |x|^{\alpha}} u(x,t) -
F\left( u(x,t) \right) = 0  ,
\ee
where $\partial^{\alpha} / \partial |x|^{\alpha}$
is the Riesz fractional derivative, and
$G_{\alpha}=g  |\Delta x|^{\alpha}$ is a finite parameter. } \\

{\bf Proof.}
To derive the equation for the field $\hat u(k,t)$, we
multiply equation (\ref{C1}) by $\exp(-ikn \Delta x)$, 
and summing over $n$ from $-\infty$ to $+\infty$. Then
\be \label{C3a}
\sum^{+\infty}_{n=-\infty} e^{-ikn \Delta x} 
\frac{\partial^s}{\partial t^s}u_n(t)=
g  \sum^{+\infty}_{n=-\infty} \
\sum^{+\infty}_{\substack{m=-\infty \\ m \not=n}}
e^{-ikn \Delta x}  J(n,m) \ [u_n-u_m] +
\sum^{+\infty}_{n=-\infty} e^{-ikn\Delta x} F(u_n) .
\ee

The left-hand side of (\ref{C3a}) gives
\be
\sum^{+\infty}_{n=-\infty} e^{-ikn \Delta x} 
\frac{\partial^s u_n(t)}{\partial t^s}=
\frac{\partial^s }{\partial t^s}
\sum^{+\infty}_{n=-\infty} e^{-ikn \Delta x} u_n(t)=
\frac{\partial^s \hat{u}(k,t)}{\partial t^s} ,
\ee
where $\hat{u}(k,t)$ is defined by (\ref{ukt}).
The second term of the right-hand side of (\ref{C3a}) is
\be
\sum^{+\infty}_{n=-\infty} e^{-ikn \Delta x} F(u_n)=
{\cal F}_{\Delta} \{F(u_n)\} .
\ee

The first term on the right-hand side of (\ref{C3a}) is
\[
\sum^{+\infty}_{n=-\infty} \ \sum^{+\infty}_{\substack{m=-\infty \\ m \not=n}} 
e^{-ikn \Delta x} J(n,m) [u_n-u_m] = \]
\be \label{C6}
=\sum^{+\infty}_{n=-\infty} \  \sum^{+\infty}_{\substack{m=-\infty \\ m \not=n}}
e^{-ikn \Delta x} J(n,m) u_n - 
\sum^{+\infty}_{n=-\infty} \sum^{+\infty}_{\substack{m=-\infty \\ m \not=n}} 
e^{-ikn \Delta x} J(n,m) u_m .
\ee
Using (\ref{ukt}) and (\ref{not}), 
the first term in r.h.s. of (\ref{C6}) gives
\be \label{C7} 
\sum^{+\infty}_{n=-\infty} \ \sum^{+\infty}_{\substack{m=-\infty \\ m \not=n}}
e^{-ikn \Delta x} J(n,m) u_n =
\sum^{+\infty}_{n=-\infty} e^{-ikn \Delta x} u_n 
\sum^{+\infty}_{\substack{m^{\prime}=-\infty \\ m^{\prime} \not=0}}
J(m^{\prime})= \hat u(k,t) \hat{J}_{\alpha}(0) ,
\ee
where we use (\ref{Jnm}) and $J(m^{\prime}+n,n)=J(m^{\prime})$, and
\be \label{not}
\hat{J}_{\alpha}(k \Delta x)=
\sum^{+\infty}_{\substack{n=-\infty \\ n\not=0}} 
e^{-ikn \Delta x} J(n)={\cal F}_{\Delta}\{ J(n)\} .
\ee
Note that
\[ \sum^{+\infty}_{n=-\infty} \ 
\sum^{+\infty}_{\substack{m=-\infty \\ m \not=n}}
e^{-ikn \Delta x} J(n,m) u_m = 
\sum^{+\infty}_{m=-\infty} u_m 
\sum^{+\infty}_{\substack{n=-\infty \\ n \not=m}} 
e^{-ikn \Delta x} J(n,m) = \]
\be \label{C9}
=\sum^{+\infty}_{m=-\infty } u_m e^{-ikm \Delta x}
\sum^{+\infty}_{\substack{n^{\prime}=-\infty \\ n^{\prime}\not=0}} 
e^{-ikn^{\prime} \Delta x} J(n^{\prime})=
\hat u(k,t)\hat{J}_{\alpha}(k \Delta x) ,
\ee
where $J(m,n^{\prime}+m)=J(n^{\prime})$ is used.

As a result, equation (\ref{C3a}) has the form
\be \label{20}
\frac{\partial^s  \hat u(k,t)}{\partial t^s}=
g [\hat{J}_{\alpha}(0)- \hat{J}_{\alpha}(k \Delta x)] \hat u(k,t) 
+{\cal F}_{\Delta} \{F(u_n)\} ,
\ee 
where ${\cal F}_{\Delta} \{F(u_n)\}$ is an operator notation for the Fourier
series transform of $F(u_n)$. \\

%%%%%%%%%%%%%%%%%%%%%%%%%%%%%%%%%%%%%%%%%%%%%%%%%%%%%%%%%%%%%%%%%%%%%%%%%

%%%{\bf Proof}. 
The Fourier series transform ${\cal F}_{\Delta}$ of (\ref{C1})
gives (\ref{20}).
We will be interested in the limit $\Delta x \rightarrow 0$. 
Using (\ref{AR}), equation (\ref{20}) can be written as
\be \label{Eq-k}
\frac{\partial^s}{\partial t^s} \hat{u}(k,t) -
G_{\alpha} \; \hat{\mathcal{T}}_{\alpha, \Delta}(k) \; \hat{u}(k,t)  
-\mathcal{F}_{\Delta} \{ F \left( u_n(t) \right) \}  = 0, 
\ee
where we use finite parameter $G_{\alpha}=g  |\Delta x|^{\alpha}$, and 
\be 
\hat{\mathcal{T}}_{\alpha, \Delta}(k) =- A_{\alpha} |k|^{\alpha}  
-R_{\alpha} (k \Delta x) |\Delta x|^{-\alpha} .
\ee
Note that $R_{\alpha}$ satisfies the condition
\[
\lim_{\Delta x \rightarrow 0} 
\frac{R_{\alpha} (k \Delta x)}{|\Delta x|^{\alpha}} =0 .
\]
The expression for $\hat{\mathcal{T}}_{\alpha,\Delta} (k)$ can be considered
as a Fourier transform of the operator (\ref{Z3b}). 
Note that $g  \rightarrow \infty$
for the limit $\Delta x \rightarrow 0$, if $G_{\alpha}$ is a finite parameter.

The passage to the limit $\Delta x \rightarrow 0$ 
for the third term of (\ref{Eq-k}) gives
\be
{\cal L}: \quad {\cal F}_{\Delta} F(u_n) \rightarrow 
{\cal L} {\cal F}_{\Delta} F(u_n) .
\ee
Then
\be
{\cal L} {\cal F}_{\Delta} \{F(u_n)\} ={\cal F} \{ {\cal L} F(u_n)\} =
{\cal F} \{ F({\cal L} u_n)\}={\cal F} \{F(u(x,t))\} ,
\ee
where we use ${\cal L} {\cal F}_{\Delta} ={\cal F} {\cal L}$. 

As a result, equation (\ref{Eq-k}) in the limit $\Delta x \rightarrow 0$ obtains
\be \label{Eq-k2}
\frac{\partial^s}{\partial t^s} \tilde{u}(k,t) - 
G_{\alpha} \; \hat{\mathcal{T}}_{\alpha}(k) \; \tilde{u}(k,t)  
-\mathcal{F} \{ F \left( u(x,t) \right) \}  = 0, 
\ee
where
\[ \tilde{u}(k,t)={\cal L} \hat{u}(k,t) , \quad 
\hat{\mathcal{T}}_{\alpha}(k) =
{\cal L}\hat{\mathcal{T}}_{\alpha, \Delta}(k) 
=-A_{\alpha} |k|^{\alpha} .
\]

The inverse Fourier transform of (\ref{Eq-k2}) gives
\be \label{Eq-x}
\frac{\partial^s}{\partial t^s} u(x,t) -
G_{\alpha} \; \mathcal{T}_{\alpha}(x) \; u(x,t) -
F\left( u(x,t) \right) = 0 ,
\ee
where $\mathcal{T}_{\alpha}(x)$ is an operator
\be \label{Tx0}
\mathcal{T}_{\alpha}(x) = 
\mathcal{F}^{-1} \{ \hat{\mathcal{T}}_{\alpha} (k) \} = 
A_{\alpha} \frac{\partial^{\alpha}}{\partial |x|^{\alpha}} .
\ee
Here, we have used the connection between the Riesz fractional 
derivative and its Fourier transform \cite{SKM}: 
$ |k|^{\alpha} \longleftrightarrow - 
{\partial^{\alpha}}/{\partial |x|^{\alpha}}$. 

As a result, we obtain continuous medium equations (\ref{CME}). $\ \ \ \Box$ \\

Examples of the interaction terms $J(n)$ that give the operators
(\ref{Tx0}) in continuous medium equations are summarized 
in the following table.
\begin{center}
\begin{tabular}{|c|c|} \hline  
& \\
$J(n)$ & $\mathcal{T}_{\alpha}(x)$  \\ & \\ \hline \hline
& \\
$ \left( \frac{(-1)^n \pi^{\alpha+1} }{\alpha+1}-
\frac{(-1)^n \pi^{1/2} }{(\alpha+1) |n|^{\alpha+1/2}}
L_1(\alpha+3/2, 1/2,\pi n) \right)$  & 
$- \partial^{\alpha}/\partial |x|^{\alpha}$ \\ 
 & \\ \hline & \\
$\frac{(-1)^n}{n^2}$ & $-(1/2) \; \partial^2 / \partial x^2 $  
\\ & \\ \hline & \\
$\frac{1}{n^2}$ & $-i \pi \; \partial / \partial x $ \\ & \\ \hline & \\
$|n|^{-(\beta+1)} , \quad (0<\beta <2, \; \beta\not=1)$ & 
$-2\Gamma(-\beta) \cos (\pi \beta/ 2) \; \partial^{\beta} / \partial |x|^{\beta} $ \\ & \\ \hline & \\
$|n|^{-(\beta+1)} , \quad (\beta >2, \; \beta\not=3,4,...)$ & 
$\zeta(\beta-1) \; \partial^2 / \partial x^2  $ \\ & \\ \hline & \\
$\frac{(-1)^n}{\Gamma(1+\alpha/2+n) \Gamma(1+\alpha/2-n)} \quad (\beta>-1/2)$ &
$-\frac{1}{\Gamma(\alpha+1)} \; \partial^{\alpha} / \partial |x|^{\alpha} $ 
\\ & \\ \hline & \\
$\frac{(-1)^n}{a^2-n^2}$ &
$-\frac{a\pi}{2\sin(\pi a)} \; \partial^2 / \partial x^2  $ \\ 
& \\ \hline & \\
$J(n)=1/n!$ & $4ei \; \partial / \partial x  $ \\ & \\ \hline
\end{tabular}
\end{center}

\newpage
%%%%%%%%%%%%%%%%%%%%%%%%%%%%%%%%%%%%%%%%%%%%%%%%%%%%%%%%%%%%%%%%%
\section{Fractional three-dimensional lattice equation}

The generalization for the three-dimensional case can be easy realized.
Let us consider the three-dimensional lattice that is described 
by the equations of motion
\be \label{3E1} 
\frac{\partial^s u_{\bf n}}{\partial t^s} =g  
\sum_{\substack{{\bf m}=-\infty \\ {\bf m} \ne {\bf n}}}^{+\infty} \; 
J({\bf n},{\bf m} ) \; [u_{\bf n} - u_{\bf m}] + F (u_{\bf n}) ,
\ee
where ${\bf n}=(n_1,n_2,n_3)$, and 
$J({\bf n},{\bf m} )=J({\bf n}-{\bf m})=J({\bf m}-{\bf n})$.
We suppose that $u_{\bf n}(t)$ are Fourier coefficients
of the function $\hat{u}({\bf k},t)$:
\be 
\hat{u}({\bf k},t) = \sum_{{\bf n}=-\infty}^{+\infty} \; 
u_{\bf n}(t) \; e^{-i {\bf k} {\bf r}_{\bf n}} =
 {\cal F}_{\Delta} \{u_{\bf n}(t)\} ,
\ee
where ${\bf k}=(k_1,k_2,k_3)$, and
\[ {\bf r}_{\bf n}=\sum^3_{i=1} n_i {\bf a}_i . \]
Here, ${\bf a}_i$ are the translational vectors of the lattice. 
The continuous medium model 
can be derived in the limit $|{\bf a}_i| \rightarrow 0$. 

To derive the equation for $\hat{u}({\bf k},t)$, 
we multiply (\ref{3E1}) by $\exp(-i {\bf k} {\bf r}_{\bf n} )$,
and summing over ${\bf n}$. Then, we obtain
\be \label{3E2}
\frac{\partial^s \hat u({\bf k},t)}{\partial t^s}=
g \left[ \hat{J}_{\alpha}(0)- \hat{J}_{\alpha}({\bf k} {\bf a}) \right] 
\hat u({\bf k},t) 
+{\cal F}_{\Delta} \{F(u_{\bf n})\} ,
\ee 
where ${\cal F}_{\Delta} \{F(u_{\bf n})\}$ is an operator notation 
for the Fourier series transform of $F(u_{\bf n})$, and
\be
\hat{J}_{\alpha}({\bf k} {\bf a})=
\sum_{{\bf n}=-\infty}^{+\infty} \; 
 e^{-i {\bf k} {\bf r}_{\bf n}} \; J({\bf n}) .
\ee

For the three-dimensional lattice, we define the $\alpha$-interaction 
with $\alpha=(\alpha_1,\alpha_2,\alpha_3)$,
as an interaction that satisfies the conditions:
\be \label{Aa2}
\lim_{k \rightarrow 0} 
\frac{[\hat{J}_{\alpha}({\bf k})- \hat{J}_{\alpha}(0)]}{|k_i|^{\alpha_i}} 
=A_{\alpha_i}, \quad (i=1,2,3) ,
\ee
where $0<|A_{\alpha_i}|< \infty$. 
Conditions (\ref{Aa2}) mean that 
\be
\hat{J}_{\alpha}({\bf k})- \hat{J}_{\alpha}(0)=
\sum^3_{i=1} A_{\alpha_i} |k_i|^{\alpha_i} +
\sum^3_{i=1} R_{\alpha_i}({\bf k}) ,
\ee
where
\be
\lim_{k_i \rightarrow 0} \ R_{\alpha_i}({\bf k}) / |k_i|^{\alpha_i}  =0 .
\ee
In the continuous limit ($|{\bf a}_i| \rightarrow 0$), 
the $\alpha$-interaction in the three-dimensional lattice 
gives the continuous medium equations with the derivatives
${\partial^{\alpha_1}}/{\partial x^{\alpha_1}}$, 
${\partial^{\alpha_2}}/{\partial y^{\alpha_2}}$, and 
${\partial^{\alpha_3}}/{\partial z^{\alpha_3}}$:
\be 
\frac{\partial^s  u({\bf r},t) }{\partial t^s}=
-g \sum^{3}_{i=1} A_{\alpha_i}
\frac{\partial^{\alpha_i} u({\bf r},t) }{\partial |x|^{\alpha_i} } 
+F(u({\bf r},t)) .
\ee 
This equation describes multifractional properties of continuous medium.

%%%%%%%%%%%%%%%%%%%%%%%%%%%%%%%%%%%%%%%%%%%%%%%%%%%%%%%%%%%%%%%%%%%%%%%%%%

\section{Linear power-law long-range interaction}

Let us consider the chain with linear long-range interaction 
that is defined by the equation of motion
\be \label{PLLRI}
\frac{\partial^s u_n}{\partial t^s} = g  
\sum_{\substack{m=-\infty \\ m \ne n}}^{+\infty} \; 
J(n,m) \; [u_n-u_m]  + F (u_n) ,
\ee
where $J(n,m)=J(|n-m|)$, and
\be \label{J(n)}
J(n)=|n|^{-(\beta+1)} 
\ee
with positive integer number $\beta$. \\

{\bf Proposition 2.}
{\it The power-law interaction (\ref{J(n)}) for the odd number $\beta$ 
is $\alpha$-interaction with $\alpha=1$ for $\beta=1$,
and $\alpha=2$ for $\beta=3,5,7...$.
For even numbers $\beta$, (\ref{J(n)}) is not $\alpha$-interaction. 
For odd number $\beta$, the transform operation $\hat T$ maps 
the equations of motion with the interaction (\ref{J(n)}) into 
the continuous medium equation (\ref{PLLRI})
with derivatives of first order for $\beta=1$,
\be \label{E1}
\frac{\partial^s}{\partial t^s} u(x,t) -
i G_{1} \; \frac{\partial}{\partial x} \; u(x,t) -
F\left( u(x,t) \right) = 0 ,
\ee
and the second order for other odd $\beta$ ($\beta=2m-1$, $m=2,3,4,...$), 
\be \label{E2}
\frac{\partial^s}{\partial t^s} u(x,t) -
G_{2} \; \frac{\partial^2}{\partial x^2} \; u(x,t) -
F\left( u(x,t) \right) = 0,
\ee
where 
\be \label{G2} 
G_1=\pi g  \Delta x, \quad
G_2=\frac{(-1)^{m-1} (2 \pi)^{2m-2}}{4(2m-2)!}  
B_{2m-2} \; g  (\Delta x)^2 \ee
are the finite parameters.  } \\

{\bf Proof}.
From Eq. (\ref{20}), we get the equation for $\hat{u}(k,t)$ in the form 
\be \label{C3b}
\frac{\partial^s \hat{u}(k,t)}{\partial t^s} + g  \; 
[\hat{J}_{\alpha}(k \Delta x)-\hat{J}_{\alpha}(0)] \; \hat{u}(k,t) - 
\mathcal{F}_{\Delta} \{ F \left( u_n(t) \right) \} =0,
\ee
where 
\be \label{C5}
\hat{J}_{\alpha}(k \Delta x) = 
\sum_{\substack{n=-\infty \\ n \ne 0}}^{+\infty} 
e^{-ikn\Delta x} |n|^{-(1+\beta)} .
\ee
The function (\ref{C5}) can be represented by
\be \label{C5ao}
\hat{J}_{\alpha}(k \Delta x)=\sum^{+\infty}_{n=1} \frac{1}{n^{1+\beta}} 
\left( e^{-ikn\Delta x} +e^{ikn\Delta x} \right) = 
2\sum^{+\infty}_{n=1} \frac{1}{n^{1+\beta}} \cos \left( k n \Delta x \right) .
\ee
Then, we can use (section 5.4.2.7 in Ref. \cite{Prudnikov}) the relations
\be
\sum^{\infty}_{n=1}\frac{\cos(n k)}{n^{2m}}=
\frac{(-1)^{m-1} (2 \pi)^{2m}}{2(2m)!}
 B_{2m} \left( \frac{k}{2\pi} \right) ,
\quad (0\le  k \le 2\pi) ,
\ee
where $m=1,2,3,...$, and $B_{2m}(z)$ are the Bernoulli polynomials \cite{BE},
which are defined by
\be
B_{n}(k)=\sum^{n}_{m=0} C^m_n B_m k^{n-m} .
\ee
Here $B_m$ are the Bernoulli numbers. 
Note $B_{2m-1}=0$ for $m=2,3,4...$ \cite{BE}.

For $\beta=1$, we have
\be
\hat{J}_{\alpha}(k \Delta x)-\hat{J}_{\alpha}(0)= 
\frac{1}{2} (k \Delta x)^2 - \pi k \Delta x  
\approx - \pi k \Delta x .
\ee
For $\beta=2m-1$ ($m=2,3,...$), 
\be
\hat{J}_{\alpha}(k)= \frac{(-1)^{m-1}}{(2m)!}
(2 \pi)^{2m} B_{2m} \left( \frac{k}{2\pi} \right) ,
\quad (0\le  k \le 2\pi) .
\ee
Then, 
\be
\hat{J}_{\alpha}(k \Delta x)-\hat{J}_{\alpha}(0)
\approx \frac{(-1)^{m-1} (2 \pi)^{2m-2}}{4(2m-2)!}  
B_{2m-2} (k \Delta x)^2 .
\ee

For $\beta=0$, we have (\cite{Prudnikov}, section 5.4.2.9.) the relation
\be
\sum^{\infty}_{n=1}\frac{\cos(n k)}{n}=-\ln \left[ 2 \sin (k/2) \right] .
\ee
Then, the limit $\Delta x \rightarrow 0$ gives
\be
\hat{J}_{\alpha}(k \Delta x) \approx - \ln (k \Delta x) 
\rightarrow \infty .
\ee
For even numbers $\beta$,
\be 
\left| \hat{J}_{\alpha}(k \Delta x)-\hat{J}_{\alpha}(0) \right| / 
\left| k \Delta x \right|^{\beta} \rightarrow  \infty 
\ee
since the expression has the logarithmic poles.

The transition to the limit $\Delta x \rightarrow 0$ 
in Eq. (\ref{C3b}) with $\beta=1$ gives 
\be \label{DD4o}
\frac{\partial^s \tilde{u}(k,t)}{\partial t^s} - 
G_1 \; k \; \tilde{u}(k,t) - 
\mathcal{F} \{ F \left( u(x,t) \right) \} =0,
\ee
where $G_1=\pi g  \Delta x$ is a finite parameter.
The inverse Fourier transform of (\ref{DD4o}) leads 
to the continuous medium equation  (\ref{E1}) with coordinate derivative 
of first order.
For $s=1$, this equation can be considered as 
the nonlinear Schr\"odinger equation.

The limit $\Delta x \rightarrow 0$ 
in Eq. (\ref{C3b}) with $\beta=2m-1$ ($m=2,3,...$) gives 
\be \label{DD5o}
\frac{\partial^s \tilde{u}(k,t)}{\partial t^s} +
G_2 \; k^2 \; \tilde{u}(k,t) - 
\mathcal{F} \{ F \left( u(x,t) \right) \} =0,
\ee
where $G_2$ is a finite parameter (\ref{G2}).
The inverse Fourier transform of (\ref{DD5o}) leads 
to the partial differential equation (\ref{E2}) of second order.
For $s=2$, this equation can be considered as a nonlinear diffusion equation,
and for $s=1$ as a nonlinear wave equation.  $\ \ \ \Box$ \\

{\bf Proposition 3.}
{\it The power-law interaction (\ref{J(n)})
with noninteger $\beta$ is $\alpha$-interaction
with $\alpha=\beta$ for $0<\beta<2$, and $\alpha=2$ for $\beta>2$.
For $0<\beta<2$ ($\beta \not=1$), 
the transform operation $\hat T$ maps the discrete equations 
with the interaction (\ref{J(n)}) into the continuous medium equation 
with fractional Riesz derivatives of order $\alpha$:
\be \label{E3}
\frac{\partial^s}{\partial t^s} u(x,t) -
G_{\alpha} A_{\alpha} \frac{\partial^{\alpha}}{\partial |x|^{\alpha}} u(x,t) =
F\left( u(x,t) \right) , \quad 
0 < \alpha < 2, \quad (\alpha \not=1) .
\ee
For $\alpha>2$ ($\alpha \not=3,4,5,...$), 
the continuous medium equation has the coordinate derivatives
of second order } \\
\be \label{E4}
\frac{\partial^s}{\partial t^s} u(x,t) +
G_{\alpha} \zeta (\alpha -1) \frac{\partial^2}{\partial |x|^2} u(x,t) =
F\left( u(x,t) \right) , \quad 
\alpha >2 , \quad (\alpha \not=3,4,...) .
\ee

{\bf Proof}.
For fractional positive $\alpha$,
the function (\ref{C5}) can be represented by
\be \label{C5b}
\hat{J}_{\alpha}(k \Delta x)=\sum^{+\infty}_{n=1} \frac{1}{n^{1+\alpha}} 
\left( e^{-ikn\Delta x} +e^{ikn\Delta x} \right) = 
Li_{1+\alpha}( e^{ik\Delta x} ) + Li_{1+\alpha}( e^{-ik\Delta x} ),
\ee
where $Li_{\beta}(z)$ is a polylogarithm function. 
Using the series representation of the polylogarithm \cite{Erd}:
\be \label{D1}
Li_{\beta}(e^z)=\Gamma(1-\beta) (-z)^{\beta-1}+\sum^{\infty}_{n=0}
\frac{\zeta(\beta-n)}{n!} z^n, \quad |z|< 2\pi, \; \; \beta\not=1,2,3...,
\ee
we obtain
\be \label{D2}
\hat{J}_{\alpha}(k \Delta x)= 
A_{\alpha} \; |\Delta x|^{\alpha} \; |k|^{\alpha} +
2 \sum^{\infty}_{n=0} 
\frac{\zeta(1+\alpha-2n)}{(2n)!} (\Delta x)^{2n} (-k^2)^n , 
\quad \alpha \not=0,1,2,3...,
\ee
where $\zeta(z)$ is the Riemann zeta-function, $|k \Delta x|< 2 \pi$, and
\be \label{D5}
A_{\alpha} =  2 \; \Gamma(-\alpha) \; 
\cos \left( \frac{\pi \alpha}{2} \right).
\ee
From (\ref{D2}), we have
\[ J_{\alpha}(0)=2 \zeta(1+\alpha) . \]
Then
\be  \label{96}
\hat{J}_{\alpha}(k \Delta x)-\hat{J}_{\alpha}(0)= 
A_{\alpha} \; |\Delta x|^{\alpha} \; |k|^{\alpha} +
2 \sum^{\infty}_{n=1} 
\frac{\zeta(1+\alpha-2n)}{(2n)!} (\Delta x)^{2n} (-k^2)^n ,  
\ee
where $\alpha \not=0,1,2,3...$, and $|k \Delta x|< 2\pi$. 

Substitution of (\ref{96}) into (\ref{C3b}) gives 
\[
\frac{\partial^s \hat{u}(k,t)}{\partial t^s} + 
g  \; A_{\alpha} |\Delta x|^{\alpha} \; |k|^{\alpha} \; \hat{u}(k,t) + \]
\be \label{D4}
+2 g  \sum^{\infty}_{n=1} \frac{\zeta(\alpha+1-2n)}{(2n)!} 
(\Delta x)^{2n} (-k^2)^n \hat{u}(k,t) -
\mathcal{F}_{\Delta} \{ F \left( u_n(t) \right) \}=0 .
\ee

We will be interested in the limit $\Delta x \rightarrow 0$. 
Then, Eq.\ (\ref{D4}) can be written as 
\be \label{Appr}
\frac{\partial^s}{\partial t^s} \hat{u}(k,t) + 
G_{\alpha} \; \hat{\mathcal{T}}_{\alpha, \Delta}(k) \; \hat{u}(k,t)  
-\mathcal{F}_{\Delta} \{ F \left( u_n(t) \right) \}  = 0, \quad 
\alpha \not=0,1,2,...,
\ee
where we use the finite parameter
\be \label{GG}
G_{\alpha}=g  |\Delta x|^{min\{\alpha;2\}} , 
\ee
and 
\be \label{D8}
\hat{\mathcal{T}}_{\alpha, \Delta}(k) = 
\begin{cases} 
A_{\alpha} |k|^{\alpha} - |\Delta x|^{2-\alpha} \zeta (\alpha -1) k^2, 
& 0 < \alpha < 2, \quad (\alpha \not=1) ;
\cr  
|\Delta x|^{\alpha-2} A_{\alpha} |k|^{\alpha} - \zeta (\alpha -1) k^2, 
& \alpha>2 , \quad (\alpha \not=3, 4, ...).
\end{cases}
\ee
The expression for $\hat{\mathcal{T}}_{\alpha,\Delta} (k)$ 
can be considered as a Fourier transform of 
the interaction operator (\ref{Z3}). 
From (\ref{GG}), we see that $g  \rightarrow \infty$
for the limit $\Delta x \rightarrow 0$, and finite value of $G_{\alpha}$.

The transition to the limit $ \Delta x \rightarrow 0$ 
in Eq. (\ref{Appr}) gives
\be \label{App2}
\frac{\partial^s}{\partial t^s} \tilde{u}(k,t)+
G_{\alpha} \hat{\mathcal{T}}_{\alpha}(k) \tilde{u} (k,t)-
{\cal F} \{ F\left( u(x,t) \right)\} =0
\quad (\alpha \not=0,1,2,...) ,
\ee
where
\be
\hat{\mathcal{T}}_{\alpha} (k) = 
\begin{cases} 
A_{\alpha} |k|^{\alpha}, & 0 < \alpha < 2, \quad \alpha \not=1; \cr 
- \zeta (\alpha -1) \; k^2, & 2< \alpha , \quad \alpha \not=3,4,....
\end{cases}
\ee

The inverse Fourier transform to (\ref{App2}) is
\be 
\frac{\partial^s}{\partial t^s} u(x,t) + 
G_{\alpha} \; \mathcal{T}_{\alpha}(x) \; u(x,t) -
F\left( u(x,t) \right) = 0 \quad 
\alpha \not=0,1,2,...,
\ee
where 
$$ \mathcal{T}_{\alpha}(x) = 
\mathcal{F}^{-1} \{ \hat{\mathcal{T}}_{\alpha} (k) \} = 
\begin{cases} 
- A_{\alpha} \; \partial^{\alpha} / \partial |x|^{\alpha} , 
& (0 < \alpha < 2, \quad \alpha \not=1); \cr 
\zeta (\alpha -1) \; \partial^2 / \partial |x|^2, 
& (\alpha >2 , \quad \alpha \not=3,4,...).
\end{cases}
$$
As the result, we obtain the continuous medium equations 
(\ref{E3}) and (\ref{E4}). $\ \ \ \Box$ \\

For $s=1$ and $F(u)=0$, Eq. (\ref{E3}) is
the fractional kinetic equation 
that describes the fractional superdiffusion \cite{Ga2,Ga5,GM2}.
If $F(u)$ is a sum of linear and cubic terms, 
then Eq. (\ref{E3}) has the form of 
the fractional Ginzburg-Landau equation 
\cite{Zaslavsky6,TZ,TZ2,Mil,Psi}.
A remarkable property of the dynamics described by the equation with 
fractional space derivatives is that the solutions have power-like tails.

%%%%%%%%%%%%%%%%%%%%%%%%%%%%%%%%%%%%%%%%%%%%%%%%%%%%%%%%%%%%%%%%%%%%%%%%%%
\section{Nonlinear long-range interaction}

Let us consider the discrete equations
with nonlinear long-range interaction:
\be \label{nli1}
\hat I_{n}(u)=\sum^{+\infty }_{\substack{m=-\infty \\ m \not= n}}
J_{\alpha}(n,m) [f(u_n)-f(u_m)] ,
\ee
where $f(u)$ is a nonlinear function of $u_n(t)$, and
$J_{\alpha}(n,m)=J_{\alpha}(n-m)$ defines the $\alpha$-interaction.
As an example of $J_{\alpha}(n)$, we can use 
\be \label{Js} 
J_{\alpha}(n)=\frac{(-1)^n}{\Gamma(1+\alpha/2+n) \Gamma(1+\alpha/2-n)} . 
\ee
For $\alpha=1,2,3,4$, the interactions with $f(u)=u^2$ and $f(u)=u-g u^2$ 
give the Burgers, Korteweg-de Vries and Boussinesq equations
in the continuous limit. 
For fractional $\alpha$ in Eq. (\ref{Js}), 
we can obtain the fractional generalization of these equations. \\

%%%%%%%%%%%%%%%%%%%%%%%%%%%%%%%%%%%%%%%%%%%%%%%%%%%%%%%%%%%%%

{\bf Proposition 4.}
{\it The transform operation maps the equations of motion 
\be  \label{nl2}
\frac{\partial^s u_n(t)}{\partial t^s}=g              
\sum^{+\infty }_{\substack{m=-\infty \\ m \not= n}}
J_{\alpha}(n-m) [f(u_n)-f(u_m)]+ F(u_n) ,
\ee
where $F$ is an external on-site force, and
$J_{\alpha}(n)$ defines the $\alpha$-interaction, into
the continuous medium equations
\be \label{nle2}
\frac{\partial^s  u(x,t)}{\partial t^s}=
G_{\alpha} A_{\alpha} \frac{\partial^{\alpha} }{\partial |x|^{\alpha}}
f(u(x,t))+F(u(x,t)) ,
\ee
where $G_{\alpha}=g |\Delta x|^{\alpha}$ is a finite parameter. } \\

{\bf Proof.}
The Fourier series transform of 
the interaction term (\ref{nli1}) can be represented as 
\[ \sum^{+\infty}_{n=-\infty} \; e^{-ikn \Delta x} 
\hat I_{n}(u)=
\sum^{+\infty}_{n=-\infty} \ 
\sum^{+\infty}_{\substack{m=-\infty \\ m \not=n}} 
e^{-ikn \Delta x} J(n,m) [f(u_n)-f(u_m)] = \]
\be \label{C6n}
=\sum^{+\infty}_{n=-\infty} \  \sum^{+\infty}_{\substack{m=-\infty \\ m \not=n}}
e^{-ikn \Delta x} J(n,m) f(u_n) - 
\sum^{+\infty}_{n=-\infty} \sum^{+\infty}_{\substack{m=-\infty \\ m \not=n}} 
e^{-ikn \Delta x} J(n,m) f(u_m) .
\ee
For the first term on the r.h.s. of (\ref{C6n}):
\be \label{C7n} 
\sum^{+\infty}_{n=-\infty} \ \sum^{+\infty}_{\substack{m=-\infty \\ m \not=n}}
e^{-ikn \Delta x} J(n,m) f(u_n) =
\sum^{+\infty}_{n=-\infty} e^{-ikn \Delta x} f(u_n) 
\sum^{+\infty}_{\substack{m^{\prime}=-\infty \\ m^{\prime} \not=0}}
J(m^{\prime})= {\cal F}_{\Delta} \{f(u_n)\} \; \hat{J}_{\alpha}(0) ,
\ee
where we use $J(m^{\prime}+n,n)=J(m^{\prime})$. 

For the second term on the r.h.s. of (\ref{C6n}):
\[\sum^{+\infty}_{n=-\infty} \ 
\sum^{+\infty}_{\substack{m=-\infty \\ m \not=n}}
e^{-ikn \Delta x} J(n,m) f(u_m) = 
\sum^{+\infty}_{m=-\infty} f(u_m) 
\sum^{+\infty}_{\substack{n=-\infty \\ n \not=m}} 
e^{-ikn \Delta x} J(n,m) = \]
\be 
=\sum^{+\infty}_{m=-\infty } f(u_m) e^{-ikm \Delta x}
\sum^{+\infty}_{\substack{n^{\prime}=-\infty \\ n^{\prime}\not=0}} 
e^{-ikn^{\prime} \Delta x} J(n^{\prime})=
{\cal F}_{\Delta} \{f(u_n)\} \; \hat{J}_{\alpha}(k \Delta x) ,
\ee
where we use $J(m,n^{\prime}+m)=J(n^{\prime})$.

As a result, we obtain
\be \label{nle1}
\frac{\partial^s  \hat u(k,t)}{\partial t^s}=
g [\hat{J}_{\alpha}(0)- \hat{J}_{\alpha}(k \Delta x)] 
{\cal F}_{\Delta} \{f(u_n)\}+{\cal F}_{\Delta} \{F(u_n)\} ,
\ee 
where $\hat{u}(k,t)={\cal F}_{\Delta}\{ u_n(t)\}$, and 
$\hat{J}_{\alpha}(k \Delta x)={\cal F}_{\Delta}\{ J(n)\}$.

For the limit $\Delta x \rightarrow 0$,  
Eq.\ (\ref{nle1}) can be written as
\be 
\frac{\partial^s}{\partial t^s} \hat{u}(k,t) -
G_{\alpha} \; \hat{\mathcal{T}}_{\alpha, \Delta}(k) \; \hat{u}(k,t)  
%%%2 g  R_{\alpha}\zeta(\alpha+1) \hat{u}(k,t) 
-\mathcal{F}_{\Delta} \{ F \left( u_n(t) \right) \}  = 0, 
\ee
where we use finite parameter $G_{\alpha}=g  |\Delta x|^{\alpha}$, and 
\be 
\hat{\mathcal{T}}_{\alpha, \Delta}(k) =- A_{\alpha} |k|^{\alpha}  
-R_{\alpha} (k \Delta x) |\Delta x|^{-\alpha} .
\ee
Here, the function $R_{\alpha}$ satisfies the condition
\[ \lim_{\Delta x \rightarrow 0} 
\frac{R_{\alpha} (k \Delta x)}{|\Delta x|^{\alpha}} =0 .\]
In the limit $\Delta x \rightarrow 0$, we get
\be \label{Eq-k3}
\frac{\partial^s}{\partial t^s} \tilde{u}(k,t) - 
G_{\alpha} \; \hat{\mathcal{T}}_{\alpha}(k) \; 
\mathcal{F} \{ f \left( u(x,t) \right) \}
-\mathcal{F} \{ F \left( u(x,t) \right) \}  = 0, 
\ee
where
\[ \tilde{u}(k,t)={\cal L} \hat{u}(k,t) , \quad 
\hat{\mathcal{T}}_{\alpha}(k) =
{\cal L}\hat{\mathcal{T}}_{\alpha, \Delta}(k) 
=-A_{\alpha} |k|^{\alpha} .
\]
The inverse Fourier transform of (\ref{Eq-k3}) gives 
the continuous medium equation (\ref{nle2}). $\ \ \ \Box$ \\

Let us consider examples of quadratic-nonlinear long-range interactions. \\

1) The continuous limit of the lattice equations
\be 
\frac{\partial u_n(t)}{\partial t}=g_{1}             
\sum^{+\infty }_{\substack{m=-\infty \\ m \not= n}}
J_1(n,m) [u^2_n-u^2_m] +
g_{2} \sum^{+\infty }_{\substack{m=-\infty \\ m \not= n}}
J_2(n,m) [u_n-u_m] ,
\ee
where $J_i(n)$ ($i=1,2$) define the $\alpha_i$-interactions
with $\alpha_1=1$ and $\alpha_2=2$, 
gives the Burgers equation \cite{Burger1} that is a nonlinear partial 
differential equation of second order:
\be
\frac{\partial}{\partial t} u(x,t)+
G_1 u(x,t)\frac{\partial}{\partial x} u(x,t)-
G_2 \frac{\partial^2}{\partial x^2} u(x,t)=0 . 
\ee
It is used in fluid dynamics as 
a simplified model for turbulence, boundary layer behavior, 
shock wave formation, and mass transport.
If we consider $J_2(n,m)$ with fractional $\alpha_2=\alpha$, then
we get the fractional Burgers equation that is suggested 
in \cite{Burger2}. In general, the fractional Burgers equation is
\be
\frac{\partial}{\partial t} u(x,t)+
G_{\alpha_1} u(x,t)\frac{\partial^{\alpha_1}}{\partial |x|^{\alpha_1}} u(x,t)-
G_{\alpha_2} \frac{\partial^{\alpha_2}}{\partial x^{\alpha_2}} u(x,t)=0 . 
\ee

2) The continuous limit of the system of equations
\be 
\frac{\partial u_n(t)}{\partial t}=g_{1}             
\sum^{+\infty }_{\substack{m=-\infty \\ m \not= n}}
J_1(n,m) [u^2_n-u^2_m] +
g_{3} \sum^{+\infty }_{\substack{m=-\infty \\ m \not= n}}
J_3(n,m) [u_n-u_m] ,
\ee
where $J_i(n)$ ($i=1,3$) define the $\alpha_i$-interactions
with $\alpha_1=1$ and $\alpha_3=3$, 
gives Korteweg-de Vries (KdV) equation 
\be
\frac{\partial}{\partial t} u(x,t)-
G_1 u(x,t)\frac{\partial}{\partial x} u(x,t)+
G_3 \frac{\partial^3}{\partial x^3} u(x,t)=0 . 
\ee
First formulated as a part of an analysis of shallow-water waves in canals, 
it has subsequently been found to be involved in a wide range of 
physics phenomena, especially those exhibiting shock waves, 
travelling waves and solitons. 
Certain theoretical physics phenomena in the quantum mechanics 
domain are explained by means of a KdV model. 
It is used in fluid dynamics, aerodynamics, and continuum 
mechanics as a model for shock wave formation, solitons, 
turbulence, boundary layer behavior and mass transport. 

If we use noninteger $\alpha_i$-interactions for $J_i(n)$,
then we get the fractional generalization of KdV equation \cite{M,M2}:
\be
\frac{\partial}{\partial t} u(x,t)-
G_{\alpha_1} u(x,t)\frac{\partial^{\alpha_1}}{\partial x^{\alpha_1}} u(x,t)+
G_{\alpha_3} \frac{\partial^{\alpha_3}}{\partial x^{\alpha_3}} u(x,t)=0 . 
\ee

3) The continuous limit of the equations
\be 
\frac{\partial^2 u_n(t)}{\partial t^2}=
g_{2} \sum^{+\infty }_{\substack{m=-\infty \\ m \not= n}}
J_2(n,m) [f(u_n)-f(u_m)] +
g_{4} \sum^{+\infty }_{\substack{m=-\infty \\ m \not= n}}
J_4(n,m) [u_n-u_m] ,
\ee
where 
\[ f(u)=u-g  u^2 , \]
and $J_i(n)$ define the $\alpha_i$-interactions
with $\alpha_2=2$ and $\alpha_4=4$, gives
the Boussinesq equation that is a nonlinear partial differential 
equation of forth order
\be
\frac{\partial^2}{\partial t^2} u(x,t)-
G_2 \frac{\partial^2}{\partial x^2} u(x,t)+
gG_2 \frac{\partial^2}{\partial x^2} u^2(x,t)+
G_4 \frac{\partial^4}{\partial x^4} u(x,t)=0 .
\ee
This equation was formulated as a part of an analysis of 
long waves in shallow water. It was subsequently applied to problems 
in the percolation of water in porous subsurface strata. 
It also crops up in the analysis of many other physical processes. 
The fractional  Boussinesq equation is
\be 
\frac{\partial^2}{\partial t^2} u(x,t)-
G_{\alpha_2} \frac{\partial^{\alpha_2}}{\partial x^{\alpha_2}} u(x,t)+
gG_{\alpha_2} \frac{\partial^{\alpha_2}}{\partial x^{\alpha_2}} u^2(x,t)+
G_{\alpha_4} \frac{\partial^{\alpha_4}}{\partial x^{\alpha_4}} u(x,t)=0 .
\ee

%%%%%%%%%%%%%%%%%%%%%%%%%%%%%%%%%%%%%%%%%%%%%%%%%%%%%%%%%%%%%%%

\section{Conclusion}

Discrete system of long-range interacting oscillators 
serve as a model for numerous applications 
in physics, chemistry, biology, etc. 
Long-range interactions are important type of interactions 
for complex media. 
An interesting situation arises when we consider 
a wide class of $\alpha$-interactions, 
where $\alpha$ is noninteger. 
A remarkable feature of these interactions is the existence
of a transform operation that replaces the set of coupled 
individual oscillator equations by the continuous medium 
equation with the space derivative of noninteger order $\alpha$.
Such a transform operation is an approximation 
that appears in the continuous limit.
This limit allows us to consider different models in unified way 
by applying tools of fractional calculus.

We can assume that an asymmetric interaction term ($J(n-m)\not=J(|n-m|)$) 
leads to other forms of the fractional derivative \cite{SKM}.

Note that a fractional derivative can be results from
a fractional difference as interaction term, just as an nth order difference
leads to an nth derivative \cite{SKM}. 
It follows from the representation of the Riesz fractional 
derivative by Grunwald-Letnikov fractional derivative:
\be 
\frac{\partial^{\alpha}}{\partial |x|^{\alpha}} u(x,t) \simeq
-\frac{1}{2 \cos(\pi \alpha /2)} 
\frac{1}{h^{\alpha}} \sum^{\infty}_{n=0} 
\frac{(-1)^n\Gamma(\alpha+1)}{\Gamma(n+1) \; \Gamma(\alpha-n+1)} \; 
\left[ u(x-nh,t) + u(x+nh,t) \right],
\ee
where $h=\Delta x$ is the discretization parameter.

A similar approach to fractional dynamics in the context of the diffusion
equation was developed in the papers \cite{R1,R2}.
In those papers, a continuum limit of (non-interacting) random particle 
motions leads to a fractional equation.

\section*{Appendix}

\begin{center}
\begin{tabular}{|c|c|} \hline  
& \\
$J(n)$ & $ \hat{J}_{\alpha}(k)-\hat{J}_{\alpha}(0)$ \\ & \\ \hline \hline
& \\
$ \left( \frac{(-1)^n \pi^{\alpha+1} }{\alpha+1}-
\frac{(-1)^n \pi^{1/2} }{(\alpha+1) |n|^{\alpha+1/2}}
L_1(\alpha+3/2, 1/2,\pi n) \right)$  & 
$ |k|^{\alpha}$ \\ 
 & \\ \hline & \\
$\frac{(-1)^n}{n^2}$ & $(1/2) \; k^2 $  \\ & \\ \hline & \\
$\frac{1}{n^2}$ & $\frac{1}{2}[k^2-2 \pi k] ,
\quad (0\le k\le 2\pi)$ \\ & \\ \hline & \\
$|n|^{-(\beta+1)} , \quad (0<\beta <2, \; \beta\not=1)$ & 
$2\Gamma(-\beta) \cos (\pi \beta/ 2) \; |k|^{\beta}$ \\ & \\ \hline & \\
$|n|^{-(\beta+1)} , \quad (\beta >2, \; \beta\not=3,4,...)$ & 
$-\zeta(\alpha-1) \; k^2 $ \\ & \\ \hline & \\
$\frac{(-1)^n}{\Gamma(1+\alpha/2+n) \Gamma(1+\alpha/2-n)} \quad (\beta>-1/2)$ &
$\frac{2^{\alpha}}{\Gamma(\alpha+1)} 
\sin^{\alpha} \left(\frac{k}{2}\right)$ \\ & \\ \hline & \\
$\frac{(-1)^n}{a^2-n^2}$ &
$\frac{\pi}{a\sin(\pi a)} \cos (ak) -\frac{1}{a^2} \quad  (0<k<2\pi)$ \\ 
& \\ \hline & \\
$J(n)=1/n!$ & $e^{\cos k} \cos ( \sin k) , 
\quad |k|<\infty $ \\ & \\ \hline
\end{tabular}
\end{center}

\end{document}